\documentclass[12pt]{iopart}

\newcommand{\vare}{\varepsilon}

\newcommand{\Tc}{$T_{\rm c}$ }
\newcommand{\Tcf}{$T_{\rm c}$}

\newcommand{\PRB}{{\it Phys. Rev. B }} 
\newcommand{\Science}{{\it Science} } 
\newcommand{\Nature}{{\it Nature} } 

\renewcommand{\r}{\vec{r}}
\renewcommand{\i}{\vec{i}}
\renewcommand{\j}{\vec{j}}
\renewcommand{\a}{\vec{a}}
\renewcommand{\b}{\vec{b}}
\newcommand{\vecdelta}{\vec{\delta}}

\newcommand{\Ce}{CeCoIn$_5$ }
\newcommand{\Cef}{CeCoIn$_5$}

\usepackage{graphicx}
\begin{document}

\title{Disordered Fulde-Ferrel-Larkin-Ovchinnikov state 
in $d$-wave superconductors} 

\author{Youichi Yanase}

\address{Department of Physics, University of Tokyo, Tokyo 113-0033, Japan}
\ead{yanase@hosi.phys.s.u-tokyo.ac.jp}
\begin{abstract}
 We study the Fulde-Ferrel-Larkin-Ovchinnikov (FFLO) superconducting state 
in the disordered systems. 
 We analyze the microscopic model, in which the $d$-wave superconductivity 
is stabilized near the antiferromagnetic quantum critical point,  
and investigate two kinds of disorder, 
namely, box disorder and point disorder, on the basis of the 
Bogoliubov-deGennes (BdG) equation. 
 The spatial structure of modulated superconducting order parameter 
and the magnetic properties in the disordered FFLO state are investigated.  
 We point out the possibility of ``FFLO glass'' state in the 
presence of strong point disorders, which arises from the 
configurational degree of freedom of FFLO nodal plane. 
 The distribution function of local spin susceptibility is calculated 
and its relation to the FFLO nodal plane is clarified. 
 We discuss the NMR measurements for \Cef. 
\end{abstract}


\section{Introduction}

 FFLO superconductivity was predicted in 1960's 
by Fulde and Ferrel~\cite{rf:Fulde1964} and also by 
Larkin and Ovchinnikov~\cite{rf:Larkin1964}. 
 In addition to the U$(1)$-gauge symmetry, a spatial symmetry 
is spontaneously broken in the FFLO state owing to the modulation of 
superconducting (SC) order parameter. 
After nearly 40 years of fruitless experimental search for FFLO states, 
recent experiments appeared to give first evidences for such 
a phase~\cite{rf:Matsudareview}.  
 Moreover, FFLO phase is attracting growing interests 
in other related fields such as the cold fermion gases~\cite{rf:Zwierlein2006} 
and the high-density quark matter~\cite{rf:Casalbuoni2004}.

 Extensive studies of FFLO state had been triggered by the discovery of 
a novel SC phase at high fields and low temperatures in the heavy fermion 
superconductor CeCoIn$_5$~\cite{rf:Radovan2003,rf:Bianchi2003}. 
 Possible FFLO states have been discovered also in some organic materials~\cite{
rf:Uji2006,rf:Singleton2000,rf:Lortz2007,rf:Shinagawa2007,rf:Yonezawa2008}. 
 All of these candidate materials are close to the antiferromagnetic 
quantum critical point (AFQCP), and then the $d$-wave superconductivity is 
expected.  
 Although it has been expected that the AFQCP significantly influences 
the superconducting state, almost all of the theoretical works on the FFLO 
state are based on the weak coupling theory and neglect 
the antiferromagnetism. 
 We have examined the FFLO state near AFQCP by analyzing the two dimensional 
Hubbard model using the FLEX approximation, and found that the $d$-wave FFLO 
state is stable in the vicinity of AFQCP owing to some strong 
coupling effects~\cite{rf:yanaseFFLO}.

 Another intriguing relationship between FFLO superconductivity and 
antiferromagnetism has been indicated in \Cef. 
 Several experimental results suggest the emergence of a FFLO state 
in \Cef~\cite{rf:Matsudareview,rf:Watanabe2004,rf:Kakuyanagi2005,
rf:Mitrovic2006,rf:Miclea2006}. 
 On the other hand, nuclear magnetic resonance (NMR) and 
neutron scattering data rather indicate the presence of 
antiferromagnetic (AFM) order in the high field phase of 
\Cef~\cite{rf:Young2007,rf:Kenzelmann2008}. 
 The pressure dependence of phase diagram~\cite{rf:Miclea2006} 
seems to be incompatible with the AFM order in the uniform SC state, 
because the AFM order is suppressed by the pressure in the other Ce-based 
heavy fermions~\cite{rf:kitaoka} while the high field phase of \Ce 
is stabilized by the pressure~\cite{rf:Miclea2006}. 
 Therefore, it is expected that the coexistent state of FFLO superconductivity 
and AFM order is realized in \Ce at ambient pressure, 
where the AFM moment is induced by the Andreev bound states around the FFLO 
nodal plane~\cite{rf:yanaseFFLOAF}.

 Another important issue of FFLO superconductivity is the role of disorders. 
 In this paper, we investigate the $d$-wave FFLO state near the AFQCP 
in the presence of randomness on the the basis of the mean field BdG equations. 
 The roles of disorder on the FFLO state has been investigated by many 
authors~\cite{rf:Aslamazov,rf:Takada,rf:Blaevskii,rf:Agterberg,rf:Adachi2003}, 
and it has been shown that the FFLO state is suppressed 
by the disorders. 
 However, the disorder average is approximately taken in these studies, 
and therefore, the regular spatial structure is artificially restored. 
 The spatial inhomogeneity is accurately taken into account using the 
BdG equations adopted in this paper. 
 We focus on the spatial structure of the disordered FFLO state 
and clarify the relationship with the magnetic properties.

 The spatial structure of $s$-wave FFLO state in the presence of 
weak box disorder has been investigated in ref.~\cite{rf:Yang}. 
 It is expected that the response to the disorder is quite different 
between the $s$-wave superconductor and $d$-wave one, because 
the $s$-wave superconductivity is robust against the disorder in accordance 
with the Anderson's theorem~\cite{rf:Anderson1959}. 
 The $d$-wave FFLO state in the presence of moderately weak point disorders 
has been investigated, and the configuration transition from two-dimensional 
structure to one-dimensional one has been pointed out~\cite{rf:Wang}. 

 In this paper, we show that the spatial structure of disordered FFLO states 
significantly depend on the feature of disorders. 
 In case of weak box disorders, the SC order parameter has 
distorted nodes, while more complicated spatial structure 
indicating the FFLO glass state is induced by the strong point disorders. 
 In the former, the magnetic properties are governed by the spatial nodes of 
SC order parameters, on which the local spin susceptibility is larger than 
that in the normal state. 
 On the other hand, the magnetic properties are dominated by 
the disorder-induced-antiferromagnetism in the latter.

 It is expected that most of our results are generally applicable to 
the FFLO state with non-$s$-wave paring. For example, the spatial structure 
of SC order parameter is independent of the details of Hamiltonian. 
 On the other hand, the disorder-induced-antiferromagnetism is a characteristic 
property of systems near AFQCP. Therefore, the magnetic properties in the 
presence of point disorders are significantly affected by the AFQCP.

 The paper is organized as follows. 
 In \S2, we formulate the BdG theory for the microscopic model 
which describes the d-wave superconductivity near AFQCP.  
 The phase diagram for the magnetic field and temperature in the clean limit 
is shown in \S3. 
 Roles of weak box disorders and strong point disorders are investigated 
in \S4 and \S5, respectively. 
 The results are summarized and some discussions are given in \S6. 
 
\section{Formulation}

 Our theoretical analysis is based on the following model 
\begin{eqnarray}
  \label{eq:model}
  && \hspace{-20mm}
  H=H_{0}+H_{\rm I} 
\\ &&\hspace{-20mm}
  H_{0} = t \sum_{<\i,\j>,\sigma} c_{\i,\sigma}^{\dag}c_{\j,\sigma}
  + t' \sum_{<<\i,\j>>,\sigma} c_{\i,\sigma}^{\dag}c_{\j,\sigma} 
  + \sum_{\i} (W_{\i}-\mu) \hspace{0.5mm} n_{\i}
  - g H \sum_{\i} S_{\i}^{\rm z} 
\\ &&\hspace{-20mm}
  H_{\rm I} = U \sum_{\i} n_{{\i},\uparrow} \hspace{0.5mm} n_{{\i},\downarrow} 
  + V \sum_{<\i,\j>} n_{\i} \hspace{1mm} n_{\j} 
  + J \sum_{<\i,\j>} \vec{S}_{\i} \hspace{1mm} \vec{S}_{\j},  
\end{eqnarray}
where $\vec{S}_{\i}$ 
is the spin operator at the site $\i$, $n_{\i,\sigma}$ is the number operator 
at site $\i$ with spin $\sigma$, and $n_{\i}=\sum_{\sigma} n_{\i\sigma}$. 
 The bracket $<\i,\j>$ and $<<\i,\j>>$ denote the summation over 
the nearest neighbour sites and next nearest neighbour sites, respectively. 
 We assume a two-dimensional square lattice. The candidate materials 
for the FFLO state, namely, CeCoIn$_5$ and organic superconductors, 
have quasi-two-dimensional Fermi surfaces. 
 We adopt the unit of energy $t=1$, and we fix $t'/t=0.25$.

 We study two kinds of disorders, which is 
taken into account in the third term of eq.~(2). 
 One is the box disorder in which the site diagonal potential $W_{\i}$ 
is randomly distributed within $[-\sqrt{3}W:\sqrt{3}W]$.  
 We multiply $\sqrt{3}$ so that the root-mean-square is 
$\bar{W} = \sqrt{<|W_{\i}|^{2}>} = W$. 
 The other is the point disorder where $W_{i}=0$ or $W_{i}=W$.
 We assume $W \ll \vare_{\rm F}$ in the former while $W \gg \vare_{\rm F}$ 
in the latter. 
 Then, the box disorder is regarded as a Born scatterer, 
while the point disorder gives rise to the unitary scattering. 
 The randomness is represented by $W$ in the former, 
while the concentration of impurity sites, where $W_{\i}=W$, determines 
the randomness in the latter. 
 The chemical potential enters in eq.~(2) as 
$\mu=\mu_{0}+\frac{1}{2} U n_{0}$,  
where $n_{0}$ is the number density at $U=V=J=H=W=0$. 
 We fix $\mu_{0}=-0.8$ for which the electron concentration is $0.8 < n < 0.9$.

 The on-site repulsive interaction is given by $U$, while 
$V$ and $J$ stand for the attractive interaction and AFM 
exchange interaction between nearest neighbour sites, respectively. 
 We take into account the AFM interaction $J$ to describe 
the FFLO state near the AFQCP. 
 The interaction $V$ stabilizes the $d$-wave superconductivity which we focus on. 
 These features, namely the $d$-wave superconductivity and AFQCP, 
can be self-consistently described using the FLEX approximation 
on the basis of the simple Hubbard model~\cite{rf:yanaseFFLO}.  
 But here, we assume the interactions $V$ and $J$ for simplicity 
in order to investigate the inhomogeneous system. 
 With the last term in eq.~(2), we include the Zeeman coupling due to the 
applied magnetic field. We assume the $g$-factor, $g=2$.

 We examine the model eq.~(1) using the BdG theory 
by taking into account the Hartree-terms arising from $U$ and $J$ 
in addition to the mean field of SC order parameter. 
 The Hartree-term due to the attractive interaction $V$ is ignored because 
this term does not have any spin dependence which is essential for the 
following results. The Hartree-term arising from $V$ may lead to the charge 
order if we assume a large attractive $V$. 
 However, we ignore this possibility since the charge ordered state is 
not stabilized in the systems near AFQCP, and 
that is an artificial consequence of the simplified model in eq.~(1).

 The mean field Hamiltonian is obtained as 
\begin{eqnarray}
  \label{eq:MFmodel}
  && \hspace{-20mm}
  H=t \sum_{<\i,\j>,\sigma} c_{\i,\sigma}^{\dag}c_{\j,\sigma}
  + t' \sum_{<<\i,\j>>,\sigma} c_{\i,\sigma}^{\dag}c_{\j,\sigma} 
  + \sum_{\i,\sigma} W_{\i,\sigma} \hspace{0.5mm} n_{\i,\sigma}, 
  - \sum_{<\i,\j>} [\Delta_{\i,\j} \hspace{1mm} c_{\i,\uparrow}^{\dag} c_{\j,\downarrow}^{\dag}
  + c.c.], 
\nonumber \\
\end{eqnarray}
where 
$W_{\i,\sigma} = W_{\i} + U <n_{\i,\bar{\sigma}}> 
+ \frac{1}{2} J \sigma \sum_{\vecdelta} <S_{\i+\vecdelta}> - H \sigma - \mu $. 
The summation of $\vecdelta$ is taken over 
$\vecdelta=(\pm 1,0), (0,\pm 1)$. 
 The pair potential is obtained as 
$\Delta_{\i,\j} = (V - J/4) <c_{\i,\uparrow} c_{\j,\downarrow}>
                - J/2 <c_{\j,\uparrow} c_{\i,\downarrow}>$ for 
$\i = \j + \vecdelta$, and otherwise $0$. 
 The thermodynamic average $<>$ is calculated on the basis of 
the mean field Hamiltonian, eq.~(4). 
 The free energy is obtained as 
\begin{eqnarray}
  \label{eq:Freeenergy}
  && \hspace{-20mm}
  F = - \sum_{\alpha} \log [1 + \exp (-E_{\alpha}/T)] 
      + \sum_{\i} W_{\i,\downarrow}
\nonumber \\ && \hspace{-10mm}
      - \frac{1}{2} \sum_{\i,\sigma} (U <n_{\i,\bar{\sigma}}> 
        + \frac{1}{2} J \sigma \sum_{\vecdelta} <S_{\i+\vecdelta}>) <n_{\i,\sigma}>
      + \sum_{\i,\j} \Delta_{\j,\i}^{\dag} <c_{\i,\uparrow} c_{\j,\downarrow}>,  
\end{eqnarray}
where $E_{\alpha}$ is the energy of Bogoliubov quasiparticles. 
 We numerically solve the mean field equations and determine the stable phase 
by comparing the free energy of self-consistent solutions. 

 The electron concentration and the magnetization at the site $\r$ is 
obtained as $n(\r)=<n_{\r,\uparrow}+n_{\r,\downarrow}>$ and 
$M(\r)=<n_{\r,\uparrow}-n_{\r,\downarrow}>$, respectively. 
 The order parameter of superconductivity is described by the pair potential 
$\Delta_{\i,\j}$. The main component of the pair potential has the $d$-wave symmetry, 
although a small extended $s$-wave component is induced in the inhomogeneous system. 
 The $d$-wave component of SC order parameter is obtained as 
\begin{eqnarray}
  && \hspace{-10mm}
\Delta^{\rm d}(\r) = \Delta_{\r,\r+\a} + \Delta_{\r,\r-\a} 
                 - \Delta_{\r,\r+\b} - \Delta_{\r,\r-\b}, 
\end{eqnarray}
where $\a = (1,0)$ and $\b = (0,1)$.

 The numerical calculation is carried out on the $N= 100 \times 100$ lattice 
in the clean limit, and on the $N = 40 \times 40$ lattice for 
disordered systems. 
 We have confirmed that qualitatively same results are obtained for 
 $100 \times 100$ and $40 \times 40$ lattices in the clean limit.

\section{Phase diagram in the clean limit}

 We first determine the phase diagram for the normal, uniform BCS, and 
FFLO states in the clean limit. We determine the stable state by 
comparing the free energy of these states. The order of phase transition 
is numerically determined by analyzing both the order parameter and free energy. 
 The free energy of two phases cross at the first order phase transition. 
 A discontinuous jump of SC order parameter also shows the first order 
transition. 
 We show that both on-site repulsion $U$ and AFM interaction $J$ 
are necessary to reproduce the phase diagram of 
CeCoIn$_5$~\cite{rf:Matsudareview}.

\begin{figure}[ht]
\includegraphics[width=16cm]{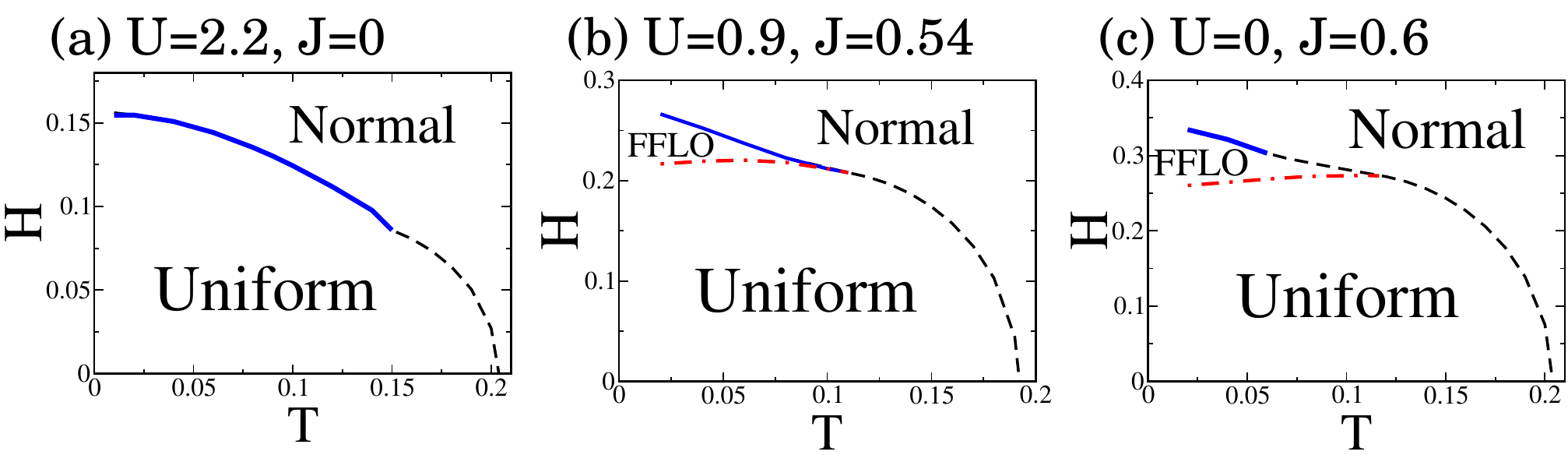}\hspace{1pc}%
\vspace*{-5mm}
\caption{Phase diagram in the clean limit ($W=0$) 
for (a) $U=2.2$ and $J=0$, (b) $U=0.9$ and $J=0.54$, and (c) $U=0$ and $J=0.6$, 
respectively. 
Blue solid lines show the first order phase transition to the SC state from the 
normal state, while black dashed lines show the second order phase transition. 
Red dash-dotted lines show the second order transition 
between the uniform BCS state and the FFLO state. 
We choose $V$ so that the transition temperature at $H=0$ 
is around $T_{\rm c} =0.2$. 
}
\end{figure}

 Figure~1 shows the phase diagram for (a) $U=2.2$ and $J=0$, 
(b) $U=0.9$ and $J=0.54$, and (c) $U=0$ and $J=0.6$. 
 For the parameters in (b), the second order phase transition occurs 
between the uniform BCS state and the FFLO state (BCS-FFLO transition). 
 The phase transition from the normal state to 
the uniform BCS state and FFLO state is first order at the temperature 
below the tricritical point, which is slightly higher than the end point of 
the BCS-FFLO transition. 
 A conventional second order superconducting transition occurs 
above the tricritical point. 
 These features of phase diagram in Fig.~1(b) are consistent with  
the experimental results for \Cef~\cite{rf:Matsudareview,
rf:Radovan2003,rf:Bianchi2003,rf:Bianchi2002,rf:Tayama2002}.

 Note that the shape of BCS-FFLO transition line seems to be
incompatible with the experimental results for \Cef. 
 A large positive slope $\partial H_{\rm BF}(T)/\partial T > 0$, where 
$H_{\rm BF}(T)$ is the magnetic field at the BCS-FFLO transition, 
has been reported in the experiments. 
 This feature does not appear in Fig.~1(b), 
however that is reproduced by taking into account 
the self-energy correction arising from the spin fluctuation near 
the AFQCP~\cite{rf:yanaseFFLO}.  
 This means that the mean field theory underestimates the stability of 
FFLO state. 
 This is not important for the spatial structure of FFLO state 
in the presence of randomness, on which we focus in this paper.

 A serious discrepancy between the theory and experiment is shown 
in the phase diagram for $J=0$ (Fig.~1(a)) and $U=0$ (Fig.~1(c)). 
 The FFLO state is completely suppressed for $J=0$, while the first order 
transition to the SC state is suppressed for $U=0$. 
 Thus, the phase diagram near the Pauli-Chandrasekhar-Clogston limit 
is significantly affected by the electron correlation. 
 These results can be understood on the basis of the Fermi liquid theory. 
 It has been have shown that the FFLO state is suppressed 
by the negative Fermi liquid parameter $F_{\rm 0a}$, while 
the first order transition to the SC state is suppressed by the 
positive $F_{\rm 0a}$~\cite{rf:Vorontsov2006}. 
 Within the mean field theory, the on-site repulsion $U$ and AFM 
interaction $J$ give rise to the negative and positive $F_{\rm 0a}$, respectively. 
 The consistency between Fig.~1(b) and experimental 
results~\cite{rf:Matsudareview,rf:Radovan2003,rf:Bianchi2003,
rf:Bianchi2002,rf:Tayama2002} indicates that the local spin fluctuation, 
which is essential for the formation of heavy fermions~\cite{rf:Yamada1986}, 
coexists with the AFM spin fluctuation in \Cef. 
 We adopt the parameters in Fig.~1(b) in the following sections.

\section{Box disorder}

 We here investigate FFLO state in the presence of box disorders, 
where the site potential $W_i$ is randomly distributed 
within $[-\sqrt{3}W:\sqrt{3}W]$. 
 Since we assume $W \ll \vare_{\rm F}$, all of the sites are weakly disordered. 

 It has been shown that a two-dimensional FFLO state can be stable 
rather than the one-dimensional FFLO state~\cite{rf:Matsudareview,rf:Wang2007}. 
 This is the case in our calculation in the clean limit ($W=0$), however 
a weak disorder ($W=0.1$) stabilizes the one-dimensional FFLO state 
as shown in Fig.~2. 
 This is qualitatively consistent with the results for moderately weak 
point disorders~\cite{rf:Wang}.

\begin{figure}[ht]
\includegraphics[width=17cm]{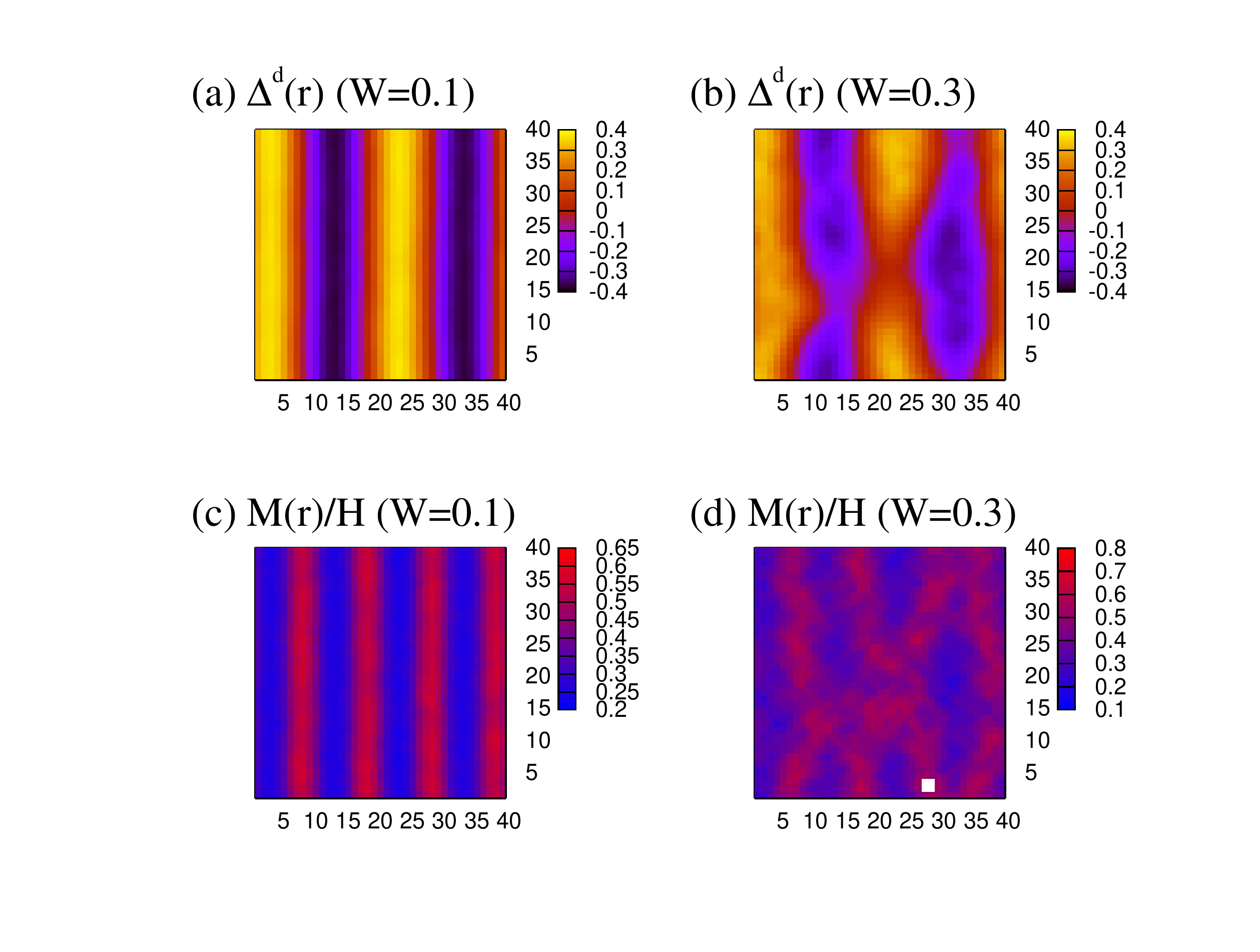}\hspace{2pc}%
\vspace*{-10mm}
\caption{(a) and (b) Typical spatial dependence of the 
$d$-wave SC order parameter $\Delta^{\rm d}(\r)$ in the presence of 
the box disorder for $W=0.1$ and $W=0.3$, respectively.
(c) and (d) Spatial dependence of the local spin susceptibility $M(\r)/H$ 
for $W=0.1$ and $W=0.3$, respectively. 
We assume $T=0.02$ and $H=0.24$ in (a) and (c), 
and adopt $T=0.02$ and $H=0.225$ in (b) and (d). 
We fix $U=0.9$, $J=0.54$, and $V=0.8$ in the following results. 
}
\end{figure}

 Figures~2(a) and (b) show a typical spatial dependence of 
the order parameter of $d$-wave superconductivity $\Delta^{\rm d}(\r)$ 
in the FFLO state for $W=0.1$ and $W=0.3$, respectively. 
 For $W=0.1$, the spatial structure of SC order parameter is almost regular, 
which is approximated by $\Delta^{\rm d}(\r) =\Delta_{0} \cos(q_{\rm f}r_{\rm x})$ 
(Fig.~2(a)). 
 On the other hand, we see a spatially modulated structure of 
SC order parameter for $W=0.3$ (Fig.~2(b)). 
 Figures~2(c) and (d) show the spatial dependence of 
local spin susceptibility $\chi(\r)=M(\r)/H$ for $W=0.1$ and $W=0.3$, 
respectively. 
 In both cases, the magnetization $M(\r)$ is induced around the 
spatial line node of SC order parameter, where $\Delta^{\rm d}(\r)=0$. 
 In particular, for a moderate disorder $W=0.3$, the spatial distribution 
of the magnetization $M(\r)$ follows the spatial nodes of SC order parameter.

\begin{figure}[ht]
\includegraphics[width=17cm]{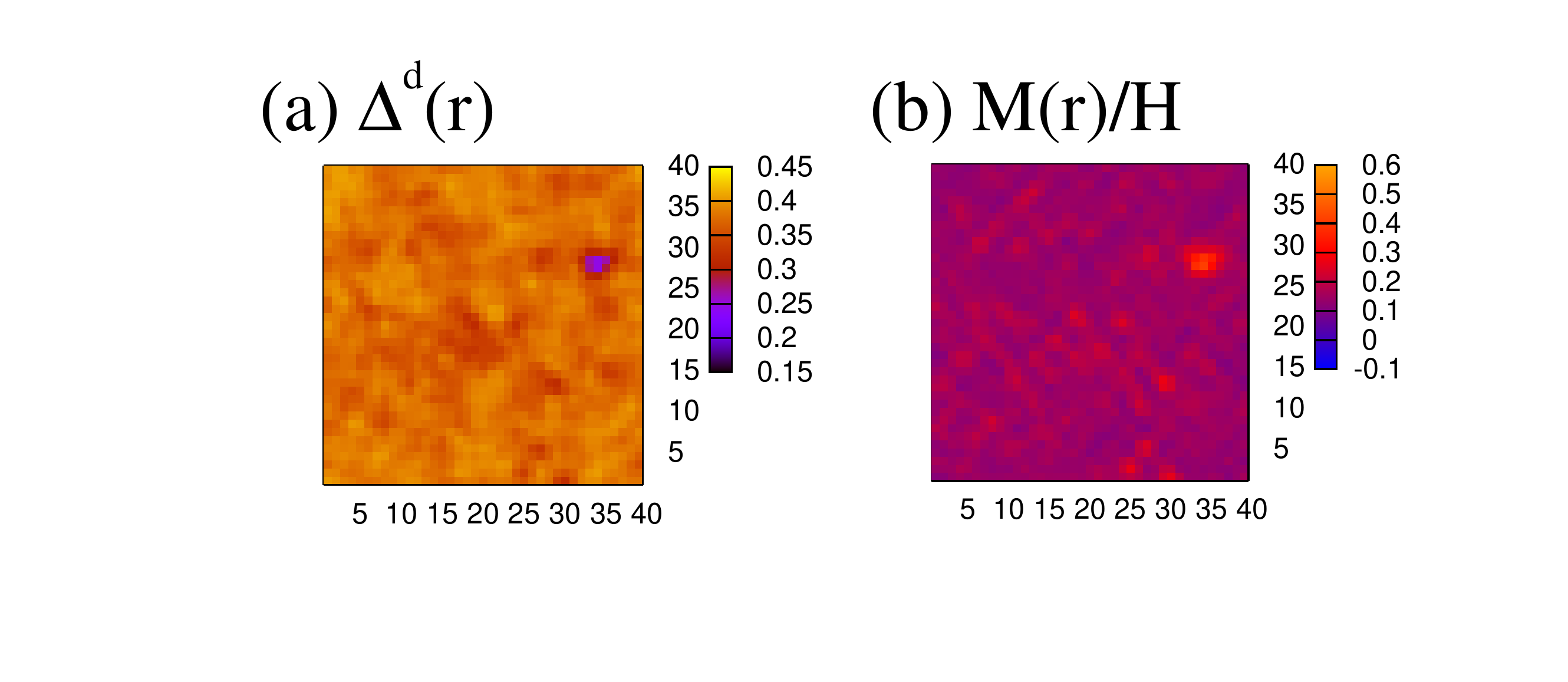}\hspace{2pc}%
\vspace*{-20mm}
\caption{(a) $D$-wave SC order parameter $\Delta^{\rm d}(\r)$ and 
(b) local spin susceptibility at $T=0.02$ and $H=0.18$. 
The disorder potential is the same as in Figs.~2(b) and (d).
}
\end{figure}

 In order to illuminate the features of FFLO state, 
we show the spatial dependences of $\Delta^{\rm d}(\r)$ and $M(\r)/H$ 
in the BCS state. 
 Fig.~3(a) shows the SC order parameter at $H=0.18$, 
where the uniform BCS state is stable in the clean limit. 
 We see that the SC order parameter is nearly uniform in the presence of 
moderately strong disorders $W=0.3$, except for the suppression 
around $\r=(35,28)$. 
 The local spin susceptibility $\chi(\r)=M(\r)/H$ is increased 
around $\r=(35,28)$ because the superconductivity is suppressed there (Fig.~3(b)). 
 We see the checkerboard structure of the local spin susceptibility, 
which is similar to high-\Tc cuprates~\cite{rf:Hoffmann2002,rf:Hanaguri2004}.  
 This checkerboard structure is induced by the quasiparticle interference 
effect~\cite{rf:Wang2003,rf:Yanase2006}. 
 The quasiparticle interference effect occurs in the FFLO state too, 
however, the spatial dependence due to the quasiparticle interference effect 
is much smaller than that arising from the inhomogeneous SC order parameter 
in the FFLO state.

\begin{figure}[ht]
\hspace*{1cm}
\includegraphics[width=14cm]{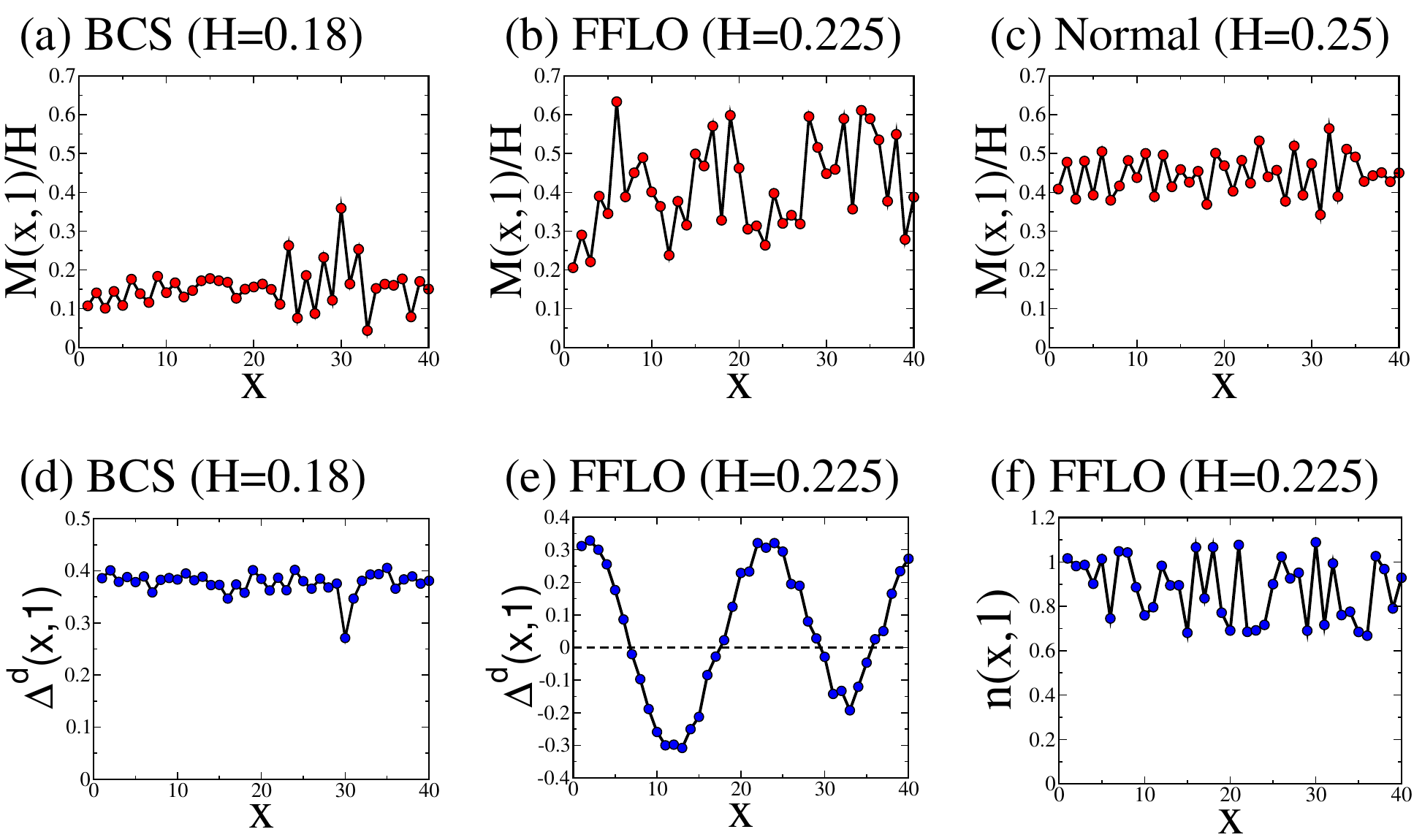}\hspace{2pc}%
\vspace*{0mm}
\caption{Spatial dependences along $\r = (x,1)$ for $W=0.3$. 
We assume the same disorder potential as in Figs.~2(b) and (d). 
Upper panel: 
Local spin susceptibility in 
(a) BCS state ($T=0.02$ and $H=0.18$), 
(b) FFLO state ($T=0.02$ and $H=0.225$), 
and (c) normal state ($T=0.2$ and $H=0.25$).  
Lower panel:
$D$-wave order parameter in (d) BCS state and (e) FFLO state. 
(f) The electron concentration $n(\r)$ in the FFLO state. 
}
\end{figure}

 To show the spatial dependences more clearly, 
 we show the local spin susceptibility, SC order parameter, and 
electron concentration along $\r = (x,1)$. 
 We see the enhancement of local spin susceptibility 
around the spatial nodes of FFLO state,  
in addition to spatial fluctuation in the atomic scale (Figs.~4(b) and (e)). 
 A large spatial dependence in the FFLO state should be contrasted to 
the small oscillation for $x < 23$ in the BCS state (Fig.~4(a)).  
 The latter arises from the quasiparticle interference effect. 
 The spatial fluctuation around $x = 30$ in the BCS state is induced by the 
inhomogeneity of SC order parameter (Fig.~4(d)). 
 The local spin susceptibility in the normal state is 
governed by the weak atomic scale oscillation (Fig.~4(c)), 
which can be regarded as a weak disorder-induced-antiferromagnetism 
(see \S4). 
 We find no clear relationship between the local spin susceptibility and 
the electron concentration in the FFLO state. The latter is shown in Fig.~4(f).

\begin{figure}[ht]
\hspace*{0cm}
\includegraphics[width=15cm]{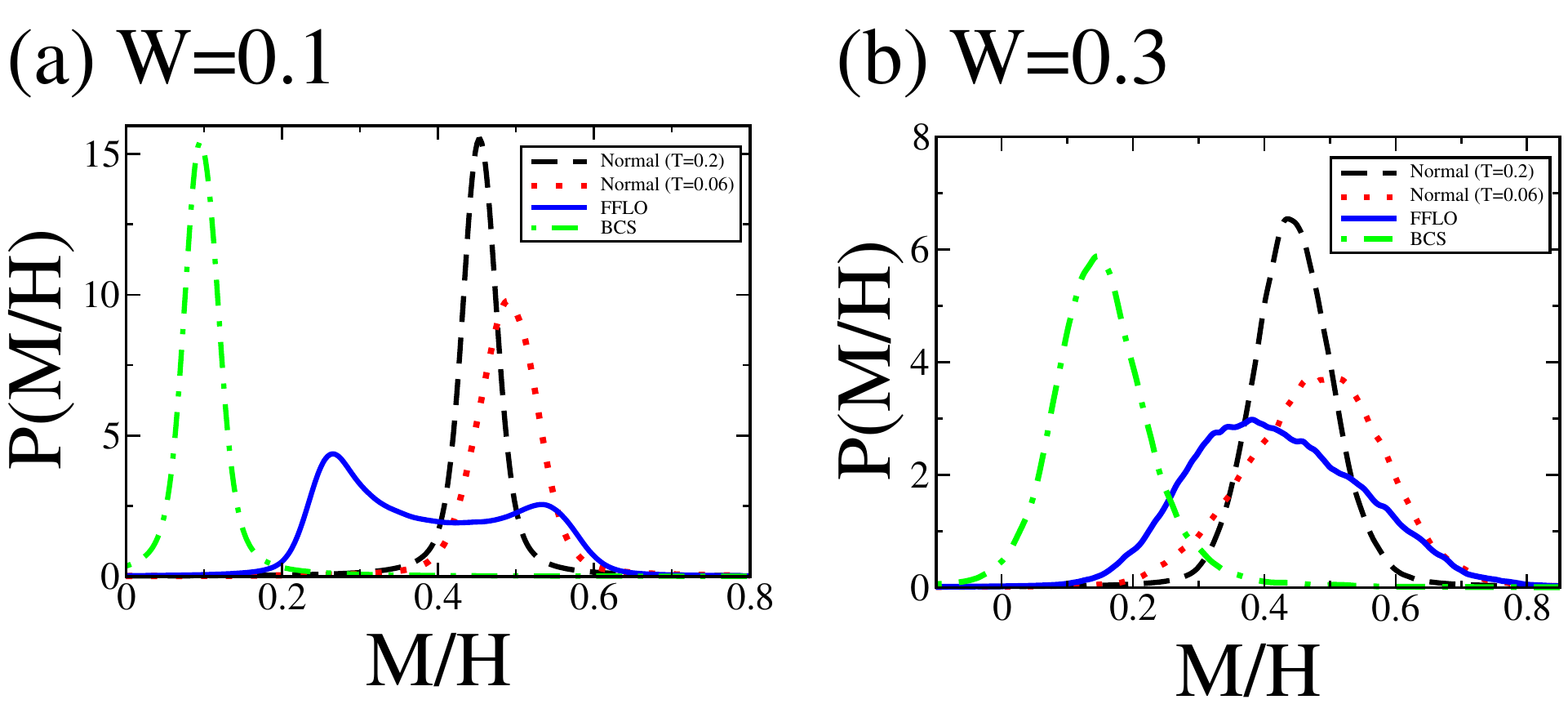}\hspace{2pc}%
\vspace*{0mm}
\caption{Distribution function of local spin susceptibility $P(M/H)$ 
for (a) $W=0.1$ and (b) $W=0.3$, respectively. 
We show the results in the BCS state ($T=0.02$ and $H=0.18$, 
green dash-dotted line), 
FFLO state ($T=0.02$ and $H=0.225$, blue solid line), 
and normal state at low temperature ($T=0.06$ and $H=0.25$, red dotted line) 
and at high temperature ($T=0.2$ and $H=0.25$, black dashed line). 
Three and five samples of the disorder potential are taken for 
the random average in (a) and (b), respectively. 
}
\end{figure}

 At the last of this section, we show the distribution function of 
local spin susceptibility $P(M/H)$, which is expressed as
\begin{eqnarray}
  && \hspace{-10mm}
P(x) = <\frac{1}{N}\sum_{\r} \delta(x-M(\r)/H)>_{\rm av}, 
\end{eqnarray}
where $<>_{\rm av}$ denotes the random average. 
 This distribution function is measured by the spectrum of NMR measurements.

 Figure~5(a) clearly shows the double peak structure of $P(M/H)$ 
in the FFLO state for a weak disorder ($W=0.1$). 
 A peak around $M/H=0.3$ arises from the region where 
the SC order parameter is large, while the other peak around $M/H=0.55$ 
comes from the Andreev bound states localized around the spatial nodes 
of SC order parameter. 
 It has been shown that this double peak structure also appears 
in the FFLO state in the presence of vortex lattice 
when the Maki parameter is large~\cite{rf:Ichioka2007}.

 As increasing the disorder potential $W$, 
the double peak structure of $P(M/H)$ in the FFLO state vanishes, 
as shown in Fig.~5(b). 
 The width of the peak in $P(M/H)$ is broader in the FFLO state 
than in the BCS state. 
 These results seem to be consistent with the NMR measurement 
of \Cef~\cite{rf:Mitrovic2006}, 
which shows a single and broad peak whose position moves to the large $M/H$ 
in the high field superconducting phase. 
 Note that the peak of $P(M/H)$ in the BCS state moves to the large $M/H$ with 
increasing the disorder potential $W$, since the residual DOS is induced by 
disorders in the $d$-wave superconductors~\cite{rf:Hirschfeld1993}. 
 This is contrasted to the FFLO state, where the average of local 
spin susceptibility $M/H$ is slightly affected by the randomness.

\section{Point disorder}

We here turn to the point disorder, in which $N = 40 \times 40$ sites 
are divided into the host sites where $W_{\i}=0$ and the impurity sites 
where $W_{\i}=W$. We assume $W=40 \gg \vare_{\rm F}$ so as to give rise to the  
unitarity scattering. 
 The impurity concentration is fixed to be $N_{\rm imp}/N=0.05$, 
where $N_{\rm imp}$ is the number of impurity sites. 
 We investigated 10 samples for the impurity distribution, 
and found that the distribution in Fig.~6(a) gives a typical result. 
 We adopt this sample in the following results.

\begin{figure}[ht]
\hspace*{-0cm}
\includegraphics[width=18cm]{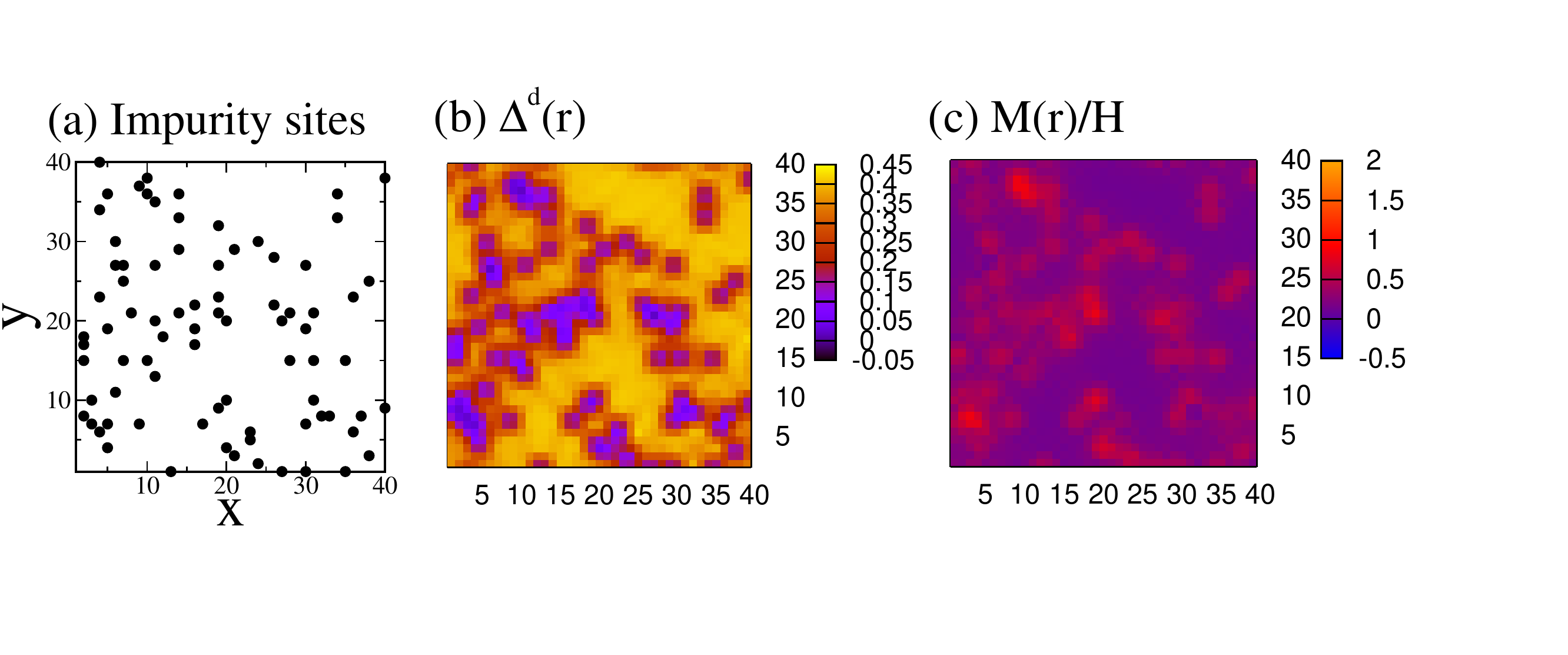}\hspace{2pc}%
\vspace*{-20mm}
\caption{(a) Typical distribution of the impurity sites. We adopt 
this sample in Figs.~6-9.  
(b) SC order parameter $\Delta^{\rm d}(\r)$ 
and (c) local spin susceptibility $M(\r)/H$ at $T=0.02$ and $H=0.18$. 
}
\end{figure}

 Figure~6(b) shows the suppression of SC order parameter around 
the impurity sites in the BCS state. 
 We see that the local spin susceptibility is significantly enhanced 
around the impurity sites (Fig.~6(c)). 
 The maximum of the local spin susceptibility is 
much larger than the spin susceptibility in the normal state 
of clean systems. 
 This is because of the disorder-induced-antiferromagnetism, 
which has been investigated in the nearly AFM Fermi 
liquid state~\cite{rf:Bulut2000}, and in the pseudogap 
state~\cite{rf:Yanase2007Proc} of high-\Tc cuprates. 
 The disorder-induced-antiferromagnetism is a ubiquitous phenomenon 
in the systems near the AFQCP, such as 
high-\Tc cuprates, organic materials, and heavy fermion systems. 
 A clear experimental evidence for the disorder-induced-antiferromagnetism 
has been obtained in high-\Tc 
cuprates~\cite{rf:Julien2000,rf:Oauzi2004,rf:Itoh2003}.

\begin{figure}[ht]
\hspace*{-2cm}
\includegraphics[width=19cm]{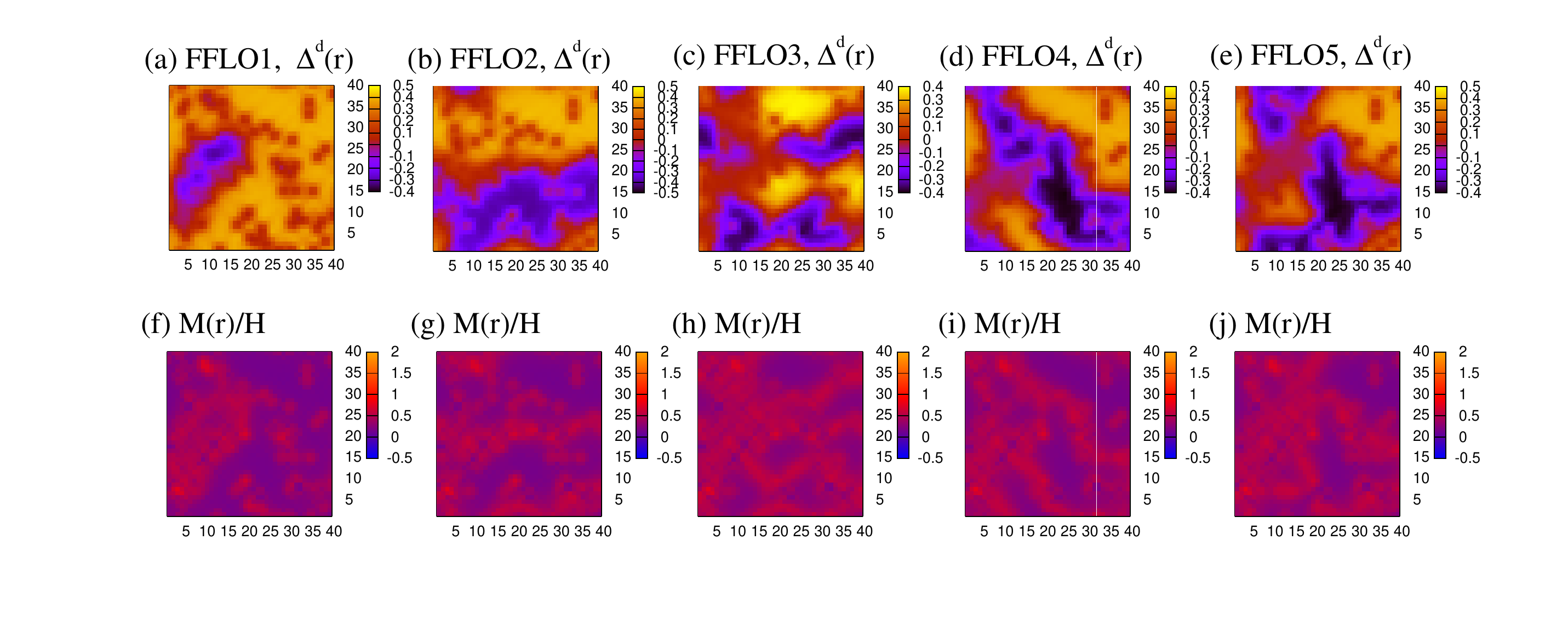}\hspace{2pc}%
\vspace*{-10mm}
\caption{Five self-consistent solutions of BdG equation at 
$T=0.02$ and $H=0.225$. 
(a-e) SC order parameter. (f-j) Local spin susceptibility. 
}
\end{figure}

 A complicated spatial structure is realized at high fields, 
where the FFLO state is stable in the clean limit. 
 Then, the free energy shows a multi-valley structure. 
 There are many local minimum of free energy with respect to 
the spatial structure of SC order parameter. 
 Figures~7(a-e) show five examples of the self-consistent solutions of 
BdG equation for the impurity distribution shown in Fig.~6(a).  
 The local spin susceptibility in each solution is shown in Figs.~7(f-j). 
 The difference of condensation energy is small between these states. 
 The condensation energy is maximum in the ``FFLO2'' state 
shown in Fig.~7(b) among the solutions obtained by us. 
 However, we obtain the solution of ``FFLO3'' state shown in Fig.~7(c) 
when we choose the SC order parameter near \Tc as a initial state of the 
mean field equation.  
 The ``FFLO1'' state in Fig.~7(a) is obtained 
when the uniform BCS state is chosen to be an initial state. 
 This means that the FFLO3 state can be stabilized as a meta-stable state 
by decreasing the temperature through \Tcf, while the FFLO1 state 
may be realized by increasing the magnetic field through the BCS-FFLO 
transition.

 The condensation energy is increased by aligning the spatial nodes of 
SC order parameter on the ``dirty region'' where the local concentration 
of impurity sites is large. 
 Many spatial configurations of the SC order parameter have 
a similar condensation energy because there are many configurations 
of spatial nodes which match the dirty region. 
 The situation is similar to the vortex glass state which is induced by the 
configurational degree of freedom of the quantum 
vortices~\cite{rf:Fisher1991,rf:Ikeda1997}. 
 The analogy with the vortex glass state indicates the possibility of 
``FFLO glass'' state, which is realized by the paramagnetic de-pairing effect 
in random systems.

 To gain the magnetic energy, the spatial node is also induced 
in the ``clean region'' at high fields.
 We found that the local spin susceptibility is enhanced around 
the spatial nodes in the clean region (Figs.~7(f-j)). 
 The quasiparticle interference effect~\cite{rf:Hoffmann2002,rf:Wang2003} 
is not visible in the presence of point disorders 
because the disorder-induced-antiferromagnetism obscures a 
weak quasiparticle interference effect.

\begin{figure}[ht]
\hspace*{2cm}
\includegraphics[width=7cm]{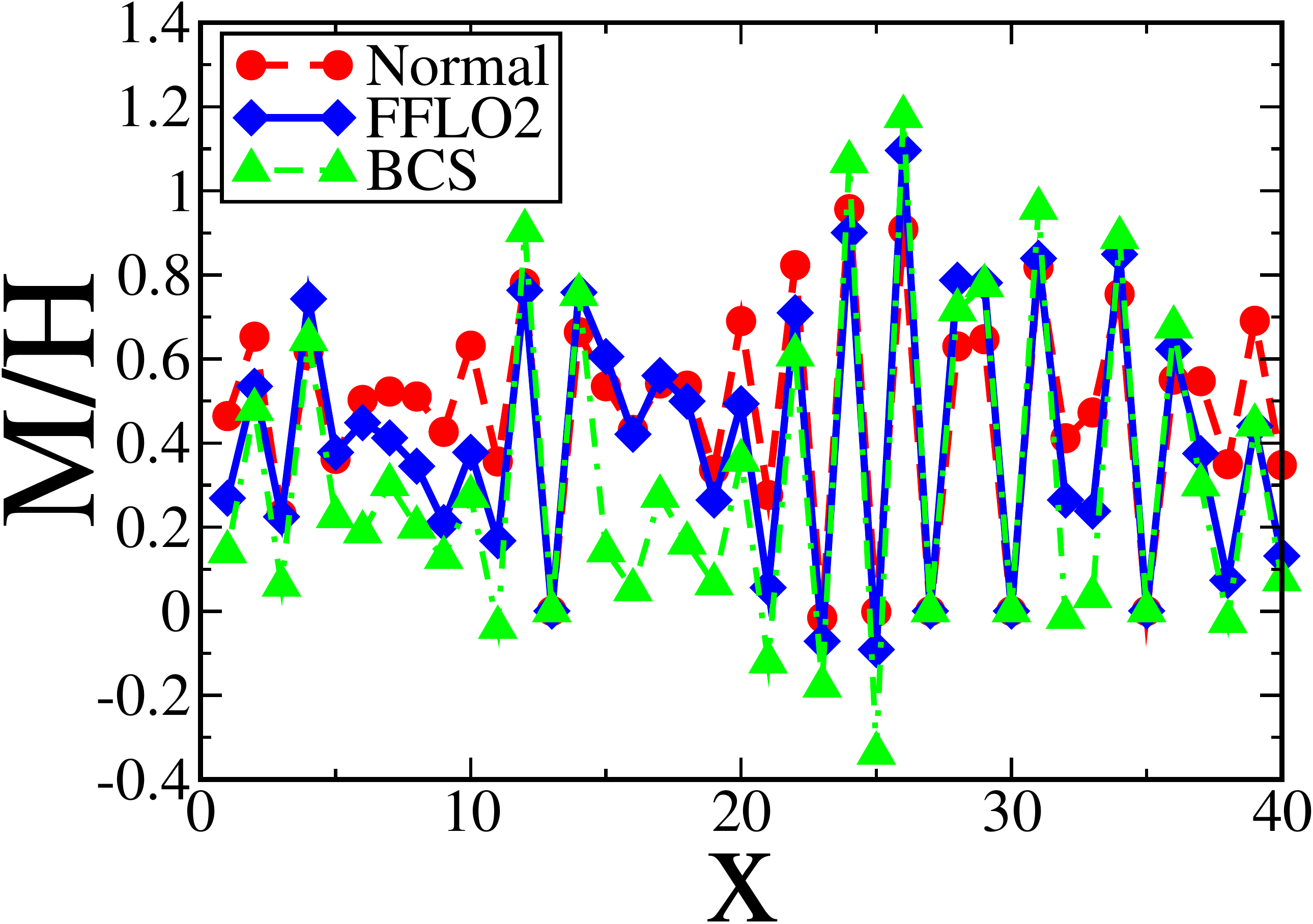}\hspace{2pc}%
\vspace*{0mm}
\caption{Local spin susceptibility along $\r = (x,1)$ in the 
BCS state (green), FFLO2 state shown in Figs.~7(b) and (g) (blue), 
and normal state (red). 
We assume $T=0.02$ and $H=0.18$ in the BCS state, 
$T=0.02$ and $H=0.225$ in the FFLO2 state, and 
$T=0.06$ and $H=0.25$ in the normal state, respectively. 
}
\end{figure}

 To clarify the local spin susceptibility arising from 
the disorder-induced-antiferromagnetism and that from the spatial nodes 
of SC order parameter, we show the $M(\r)/H$ along $\r = (x,1)$ in Fig.~8. 
 The normal state, FFLO2 state, and BCS state are shown for a comparison. 
 As shown in Fig.~6(a), the system is clean at $x < 20$, while it is dirty 
at $x >20$. 
 Figure~8 shows that the disorder-induced-antiferromagnetism occurs and 
gives rise to the significant oscillation of the magnetization 
in the dirty region ($x > 20$). 
 This is a ubiquitous phenomenon near the AFQCP in the absence of 
translational symmetry. 
 We see that the disorder-induced-antiferromagnetism is enhanced in the 
FFLO state as well as in the BCS state. 
 The local spin susceptibility is more significantly affected by the 
superconductivity in the clean region ($x < 20$). 
 It is shown that the spin susceptibility is larger in the FFLO2 state 
than in the BCS state because of the presence of spatial node around 
$x=15$ (Fig.~7(b)).

\begin{figure}[ht]
\hspace*{2cm}
\includegraphics[width=7cm]{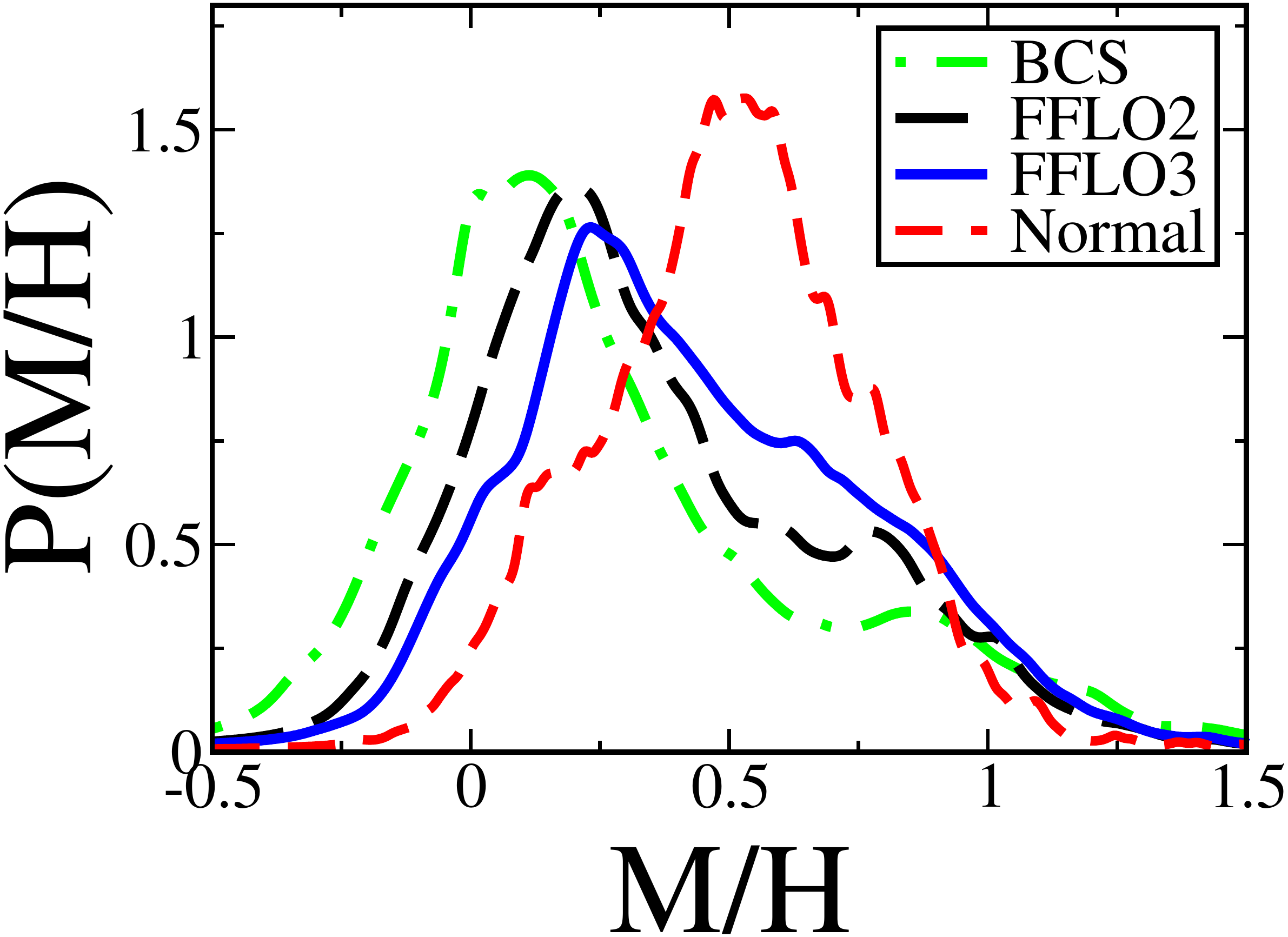}\hspace{2pc}%
\vspace*{0mm}
\caption{Distribution function of the local spin susceptibility $P(M/H)$ 
in the presence of $5\%$ point disorders. 
The BCS state (green dash-dotted line), 
FFLO2 state in Figs.~7(b) and (g) (black dashed line), 
FFLO3 state in Figs.~7(c) and (h) (blue solid line), 
and normal state (red dashed line) are shown for a comparison.  
The parameters $H$ and $T$ in each state are the same as in Figs.~7 and 8. 
}
\end{figure}

 Figure~9 shows the distribution function of the local spin susceptibility 
$P(M/H)$, which is defined as 
\begin{eqnarray}
  && \hspace{-10mm}
P(x) = <\frac{1}{N}\sum_{\r}^{{\rm host}} \delta(x-M(\r)/H)>_{\rm av}.  
\end{eqnarray}
 The summation $\sum_{\r}^{{\rm host}}$ is taken over the host sites and 
therefore the contribution form the impurity sites is eliminated in eq.~(8). 
 The double peak structure appears in the BCS state as well as 
in the FFLO2 state because of the the disorder-induced-antiferromagnetism.  
 This structure vanishes in the FFLO3 state in which 
many spatial nodes exists in the SC order parameter. 
 These results are incompatible with the NMR measurements 
for \Cef~\cite{rf:Mitrovic2006,rf:Young2007}. 
 The width of the peak hardly changes through the BCS-FFLO transition 
in contrast to ref.~\cite{rf:Mitrovic2006}. 
 This means that the model based on the point disorder is not 
relevant for \Cef. 
 However, the point disorders can be systematically induced by substituting 
Ce ions by La ions, or In ions by Cd ions. 
 Therefore, it is interesting to investigate the superconducting state 
in Ce$_{1-x}$La$_x$CoIn$_5$ and CeCoIn$_{5-x}$Cd$_{x}$~\cite{rf:Tokiwa} 
at high fields.

\begin{figure}[ht]
\hspace*{-2cm}
\includegraphics[width=15cm]{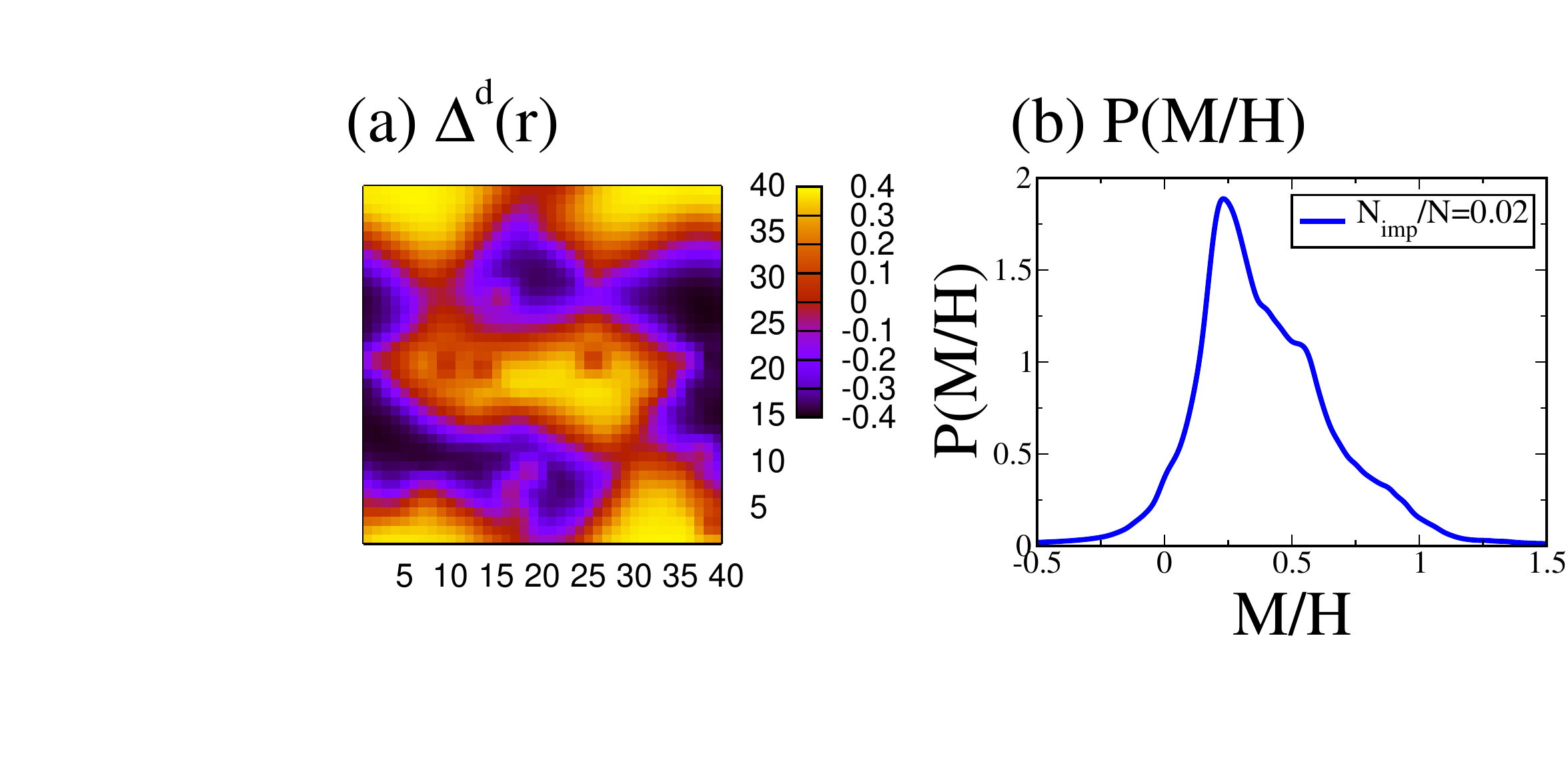}\hspace{2pc}%
\vspace*{-12mm}
\caption{(a) Typical spatial dependence of SC order parameter  
and (b) distribution function of local spin susceptibility 
in the presence of $2\%$ point disorders. 
We assume $T=0.02$ and $H=0.235$. 
}
\end{figure}

 Before closing this section, we briefly discuss the case of dilute impurities. 
 Figure~10(b) shows the distribution function of the local spin susceptibility 
in the presence of $2\%$ point disorders. 
 We show the result for a typical distribution of impurities  
which leads to the SC order parameter shown in Figure~10(a). 
 We see the structures around $M/H=0.5$ and $M/H=0.8$ in addition to the 
pronounced peak at $M/H=0.25$. 
 The plateau around $M/H=0.5$ is owing to the spatial nodes of 
SC order parameter, while the structure around $M/H=0.8$ arises from 
the disorder-induced-antiferromagnetism.

\section{Summary and discussion}

 We investigated the disordered FFLO state with focus on the spatial structure of 
SC order parameter and the magnetic properties. 
 In particular, the $d$-wave superconductivity near the AFQCP 
has been investigated in details. 
 It has been shown that the spatial dependence of SC order parameter is 
relatively simple in case of the box disorder; 
the FFLO nodal plane is modulated. 
 On the other hand, the spatial structure is complicated in the presence of  
point disorders. Then, the spatial nodes are strongly pinned to the 
locally dirty region. There are many configurations in which the nodes are 
pinned to the point disorders, and therefore a glassy behaviour appears.  
 We called this state ``FFLO glass '' in analogy with the vortex glass state. 

 We have shown that the magnetic properties in the SC state significantly 
depend on the character of disorders. 
 In the FFLO state with box disorders, the magnetic properties are 
governed by the nodal plane of SC order parameters, on which the local spin 
susceptibility is larger than that in the normal state. 
 On the other hand, the magnetic properties in the BCS state are 
mainly determined by the quasiparticle interference effect, 
which gives rise to an oscillation with a small amplitude. 
 A weak disorder-induced-antiferromagnetism is induced by the inhomogeneity 
of SC order parameters in the BCS state as well as in the FFLO state.

 The magnetic properties are dominated by the disorder-induced-antiferromagnetism 
in the presence of point disorders. 
 We found that the disorder-induced-antiferromagnetism is enhanced by the 
superconductivity in both BCS and FFLO states. 
 The local spin susceptibility in the locally clean region is suppressed 
in the BCS state, while that is increased in the FFLO state owing to the 
spatial nodes of SC order parameter.

 Finally, we examine the distribution function of local spin susceptibility 
$P(M/H)$ and discuss the NMR measurements for \Cef. 
 It has been known that the distribution function shows a double peak structure 
in the FFLO state in the clean limit~\cite{rf:Ichioka2007}. 
 We have shown that the two peaks merge into the single peak in the 
presence of box disorders and/or point disorders. 
 However, the distribution function is affected in a different way 
by the box disorder and point disorder. 
 The single peak structure is induced by the box disorders 
because of the displacement of FFLO nodes, while 
the disorder-induced-antiferromagnetism is the main cause of the broad 
single peak in the presence of point disorders. 
 We can distinguish these two cases by analyzing the line width of $P(M/H)$. 
 The line shape of $P(M/H)$ is very broad and its width does not change 
through the BCS-FFLO transition in case of the point disorder (Fig.~9). 
 This is because the local antiferromagnetism occurs near the point 
disorders in both BCS and FFLO states (Fig.~8). 
 On the other hand, the line shape is significantly broadened 
in the FFLO state through the BCS-FFLO transition 
in case of the box disorders, as shown in Fig.~5. 
 The latter seems to be consistent with the NMR measurement for \Ce at the 
In(1) site~\cite{rf:Mitrovic2006}.

 Another NMR spectrum at the In(2) site has shown the double peak 
structure with large splitting and indicated the AFM order 
in the high field phase of \Ce~\cite{rf:Young2007}. 
 A clear experimental evidence for the AFM order has been obtained 
by the recent neutron scattering measurement~\cite{rf:Kenzelmann2008}. 
 We have shown that the AFM order can occur in the FFLO state near the AFQCP 
and the phase diagram suggested in refs.~\cite{rf:Young2007,rf:Kenzelmann2008} 
is consistent with the coexistence of antiferromagnetism 
and FFLO superconductivity~\cite{rf:yanaseFFLOAF}. 
 Although the double peak structure of the NMR line shape is also induced 
by the disorder-induced-antiferromagnetism (Fig.~9), 
this is not the case in \Cef. 
 The direction of the AFM moment is parallel to 
the applied magnetic field in case of the disorder-induced-antiferromagnetism. 
 However, the neutron scattering measurement shows the magnetic moment 
perpendicular to the magnetic field~\cite{rf:Kenzelmann2008}, which 
means the spontaneous symmetry breaking. 
 Therefore, the true long range order of antiferromagnetism seems to occur 
in the experiments of refs.~\cite{rf:Young2007,rf:Kenzelmann2008}.

 Our results in this paper suggest that the single peak structure 
of NMR spectrum at In(1) site, which is less sensitive to the AFM order 
than In(2) site, can be caused by the disorder. 
 But, the AFM order may be another cause. 
 We are planning to investigate the magnetic properties in the 
coexistent state of AFM order and FFLO superconductivity. 
 From the experimental point of view, it would be interesting to 
investigate the pressure effect in \Cef. 
 It is expected that the AFM order is suppressed by the pressure 
while the FFLO superconductivity is enhanced~\cite{rf:yanaseFFLO}. 
 The latter has been observed in ref.~\cite{rf:Miclea2006}. 
 Therefore, the pure FFLO state may be realized at high pressures.

 According to these considerations, the point disorder is not 
the main source of the randomness in \Cef. However, the point disorder 
can be induced by substituting Ce ions with La ions or In ions with Cd ions. 
 Thus, Ce$_{1-x}$La$_x$CoIn$_5$ and CeCoIn$_{5-x}$Cd$_{x}$~\cite{rf:Tokiwa} 
will be an intriguing playground of the FFLO superconductivity 
with point disorders.

 In summary, we investigated the FFLO superconducting state 
in the presence of randomness. 
 The spatial structure of SC order parameter and 
magnetic properties are clarified in details. 
 It has been proposed that the experimental results on \Ce can be understood 
by assuming the FFLO state realized in the high field phase 
and taking into account both the AFM spin correlation and a weak disorder.

\section*{Acknowledgments}
 The authors are grateful to M. Ichioka, M. Kenzelmann, K. Kumagai, 
K. Machida, Y. Matsuda, V. F. Mitrovic, and M. Sigrist  
for fruitful discussions. 
 This study has been financially supported by 
Grants-in-Aid for Scientific Research on Priority Areas
"Physics of New Quantum Phases in Superclean Materials" and 
for Young Scientists (B) from the MEXT. 
 Numerical computation in this work was carried out 
at the Yukawa Institute Computer Facility.

\end{document}